\documentclass[a4paper, 11pt]{article}
\usepackage{jheppub} 
\usepackage{jheppub} 
\usepackage[utf8]{inputenc}
\usepackage{physics}
\usepackage{slashed}
\usepackage{cancel}
\usepackage{caption}
\usepackage{xcolor}
\usepackage{comment}
\usepackage{multirow}
\usepackage{graphics}
\usepackage{float}
\usepackage{cancel}
\usepackage{soul}
\usepackage{cases}
\usepackage{array}
\usepackage{mathtools}  
\usepackage{amsfonts}
\usepackage{hyperref}
\usepackage{amsmath}
\usepackage{amssymb}
\usepackage{tcolorbox}
\usepackage{tikz}
\usepackage{tikz-feynman}
\tikzfeynmanset{compat=1.1.0}

\title{ Super-Grassmannians for $\mathcal{N}=2$ to $4$ SCFT$_3$: From AdS$_4$ Correlators to $\mathcal{N}=4$ SYM scattering Amplitudes}
\author{Aswini Bala,  Sachin Jain,  Dhruva K.S.,  Adithya A Rao}

\affiliation{Indian Institute of Science Education and Research, \\ Dr Homi Bhabha Road,  Pashan,  Pune,  India}

\emailAdd{aswini.bala@students.iiserpune.ac.in}
\emailAdd{sachin.jain@iiserpune.ac.in}
\emailAdd{k.s.dhruva@students.iiserpune.ac.in}
\emailAdd{adithya.arao@students.iiserpune.ac.in}

\abstract{We construct a Super-Grassmannian for $n-$point functions in $\mathcal{N}=2$ to $4$ SCFT$_3$. The constraints imposed by super-conformal invariance and $R-$symmetry are completely manifest in this formalism through (operator-valued) delta functions. We test our formalism in $\mathcal{N}=2$ and $\mathcal{N}=4$ AdS$_4$ super Yang–Mills theories. In the $\mathcal{N}=2$ case, for instance, we reproduce the four-gluon amplitude using the four-point scalar correlator as input.  For $\mathcal{N}=4$, we construct the super-operator in two distinct ways. In one approach, the super-operator has a lowest component of spin zero and includes all states up to spin two. In the other approach, we build the super-operator in a CPT self-conjugate manner, which contains only operators with spin zero, spin half, and spin one mimicking flat space  $\mathcal{N}=4$ SYM super-field constructions. The latter construction is particularly interesting, as it matches directly with the $\mathcal{N}=4$ SYM amplitudes in the flat space limit, thereby demonstrating the non-triviality and usefulness of our framework. It is interesting to note that the $R-$symmetry group enhances from $SO(\mathcal{N})$ to $SU(\mathcal{N})$ in the flat space limit.}
\emailAdd{~}

\begin{document}

\maketitle

\section{Introduction}
Supersymmetry plays a central role in simplifying and organizing scattering amplitudes. By relating particles of different spins within the same supermultiplet, SUSY unifies seemingly distinct amplitudes into a single superamplitude, thereby drastically reducing the number of independent quantities that need to be computed \cite{Elvang:2013cua}. This leads to powerful constraints from Ward identities, often fixing entire classes of amplitudes or determining them in terms of a few basic building blocks. In highly supersymmetric theories like $\mathcal{N}=4$ SYM, these constraints become so powerful that scattering amplitudes are remarkably simple. They often exhibit hidden symmetries, such as dual conformal invariance, and admit very compact representations \cite{Drummond:2008vq}. Supersymmetry, together with on-shell methods like recursion relations and Grassmannian constructions, provides an efficient and direct way to build amplitudes from basic principles \cite{Arkani-Hamed:2009hub, Arkani_Hamed:duality}. Altogether, this makes it a powerful framework for revealing the underlying structure of scattering processes in quantum field theory.

Building on these ideas, it is natural to ask whether the success of scattering amplitudes can be replicated in conformal field theories, particularly in three dimensions. Recent developments have shown that many amplitude-inspired ideas can be translated into the language of correlation functions in CFT such as using spinor helicity \cite{Maldacena:2011nz, McFadden:2011kk,  Coriano:2013jba,  Bzowski:2013sza,  Ghosh:2014kba,  Bzowski:2015pba,  Bzowski:2017poo,  Bzowski:2018fql,  Farrow:2018yni,  Isono:2019ihz,  Bautista:2019qxj,  Gillioz:2019lgs,  Baumann:2019oyu,  Baumann:2020dch,  Jain:2020rmw,  Jain:2020puw,  Jain:2021wyn,  Jain:2021qcl,  Jain:2021vrv,  Baumann:2021fxj,  Jain:2021gwa,  Jain:2021whr,  Gillioz:2022yze,  Marotta:2022jrp,  Jain:2023idr,  Bzowski:2023jwt,  S:2024zqp,  Jain:2024bza,  Aharony:2024nqs,  Marotta:2024sce,  Coriano:2024ssu,  Gillioz:2025yfb,  S:2025pmh}, twistors \cite{Baumann:2024ttn,  Bala:2025gmz,  Bala:2025jbh,  Bala:2025qxr,  Rost:2025uyj,  S:2025pmh,  Mazumdar:2025egx,  CarrilloGonzalez:2025qjk, Ansari:2025fvi} and the most recent Grassmannian framework \cite{Arundine:2026fbr}. In this context, Grassmannian formulations provide a promising framework to encode symmetries, kinematics, and supersymmetry in a unified way. In this work, we extend these ideas to $\mathcal{N}=2$ to $\mathcal{N}=4$ supersymmetric theories in three-dimensional CFTs. 

An interesting testing ground for these ideas is provided by AdS, where the CFT formalism developed above can be directly applied to compute observables. In this setting, AdS amplitudes—computed via Witten diagrams—play a role analogous to scattering amplitudes in flat space. This makes it possible to ask whether the same structural simplicity and organizing principles, such as recursion, factorization, and geometric formulations, persist in a curved background. Importantly, once these techniques are formulated in the CFT language, they can be readily used to access AdS observables, allowing us to systematically explore how amplitude-inspired ideas extend beyond flat space.

In spinor helicity variables, supersymmetry relates component correlators through a combination of algebraic relations and first-order differential equations. While these constraints are powerful, they can be technically involved to implement in practice. In contrast, the Grassmannian formulation leads purely to algebraic relations among correlators, making the structure significantly more transparent and easier to use \cite{Bala:2026new1}. This simplification motivates us to ask the following key questions:
\begin{enumerate}
    \item Can the Grassmannian framework be used to efficiently reconstruct higher-spin correlators—such as the four-graviton amplitude—starting from simpler inputs like the four-point scalar correlator? While such observables could, in principle, be obtained directly through bootstrap methods, verifying the correctness of the result is often nontrivial in spinor helicity variables. In contrast, scalar correlators are much easier to compute and check, making them a natural starting point.
    
    \item Upon taking an appropriate flat-space limit (for instance, in the $\mathcal{N}=4$ case), does the Grassmannian construction reproduce the known flat-space $\mathcal{N}=4$ SYM amplitudes?
\end{enumerate}

These questions highlight both the practical advantages and the conceptual reach of the Grassmannian approach.

This work builds on and generalizes our companion paper, where the construction of a Super-Grassmannian description was developed for $\mathcal{N}=1$. Here, we extend the framework to higher supersymmetry and formulate a Super-Grassmannian description for $n$-point functions in $\mathcal{N}=2,3,4$ SCFT$_3$. A key feature of our approach is that super-conformal symmetry, along with the associated $R$-symmetry, is implemented in a manifest way, leading to a compact and unified encoding of the relevant constraints. We illustrate the power of this formalism through explicit examples in AdS$_4$ super Yang–Mills theories. In particular, for $\mathcal{N}=2$, we show how nontrivial higher-spin observables can be reconstructed starting from scalar correlators. For $\mathcal{N}=4$, we explore two different constructions of the super-operator, one of which closely parallels the structure familiar from flat-space super-fields. Remarkably, in an appropriate flat-space limit, this construction reproduces known $\mathcal{N}=4$ SYM amplitudes, providing a nontrivial consistency check of our framework.

\subsection*{Outline}
In section \ref{sec:Formalism}, we develop the formalism for generic $n-$point super-correlators in $\mathcal{N}-$extended supersymmetry. We specialize to $\mathcal{N}=2$ SCFTs in section \ref{sec:Formalism} and discuss explicitly the construction of two, three, and four-point correlators. We then apply this to the setting of $\mathcal{N}=2$ AdS$_4$ SYM theory in section \ref{sec:Neq2SYM}. In section \ref{Neq4SYM}, we discuss $\mathcal{N}=4$ SYM theory as another important application of our formalism. Finally, we take the flat space limit of our results and match with the existing results in section \ref{sec:FlatLimit}. We conclude with a summary and potential future directions in section \ref{sec:Discussion}. Further, in appendix \ref{app:Notation}, we state our basic conventions and notations. \\
\textbf{Note}:~\textit{We have made an attempt to make this paper entirely self-contained, although the reader may wish to consult our companion $\mathcal{N}=1$ work \cite{Bala:2026new1}.}

\section{ The $\mathcal{N}=2,3,4$ Orthogonal Super-Grassmannian $\mathcal{OG}r(n,2n)$}\label{sec:Formalism}
The aim of this section is to construct the orthogonal super-Grassmannian for $\mathcal{N}=2,3,4$ SCFT$_3$, generalizing the result of our companion $\mathcal{N}=1$ paper to higher supersymmetry. We begin by discussing the representation of the super-conformal algebra acting on our super-fields, followed by setting up the super-Grassmannian that automatically solves the super-conformal Ward identities.

\subsection{The Super-Conformal Generators and Ward Identities}
The arena we work is the super-space spanned by the spinor helicity and Grassmann twistor coordinates $(\lambda^a,\Bar{\lambda}^a,\xi^\alpha)$ or $(\lambda^a,\Bar{\lambda}^a,\Bar{\xi}^\alpha)$ where $a=1,2$ are $\mathfrak{sl}(2)$ spinor indices and $\alpha=1,\cdots,\mathcal{N}$ is a vector index of the $R-$symmetry group $SO(\mathcal{N})$. $\Bar{\xi}^\alpha$ and $\xi^\alpha$ are Grassmann Fourier conjugate to each other.

The action of the supersymmetry generator acting on super-currents of interest to us in these variables is\footnote{We will be more explicit about the exact component expansion of the super-currents subsequently.},  
\begin{align} \label{Qgenerator}
    \mathcal{Q}^\alpha_a(\xi)=\frac{1}{\sqrt{2}}\bigg(\Bar{\lambda}_a \xi^\alpha+\lambda_a\frac{\partial}{\partial \xi_\alpha}\bigg),  ~\mathcal{Q}^\alpha_a(\Bar{\xi})=\frac{1}{\sqrt{2}}\bigg(\Bar{\lambda}_a\frac{\partial}{\partial\Bar{\xi}_\alpha}+\lambda_a \Bar{\xi}^\alpha\bigg).
\end{align}
The special super-conformal generator,   on the other hand,   takes the form,
\begin{align}\label{Sgen}
    \mathcal{S}^\alpha_a(\xi)=\frac{1}{\sqrt{2}}\bigg(\xi^\alpha\frac{\partial}{\partial\lambda^a}+\frac{\partial}{\partial\xi_\alpha}\frac{\partial}{\partial\Bar{\lambda}^a}\bigg),  ~ \mathcal{S}^\alpha_a(\Bar{\xi})=\frac{1}{\sqrt{2}}\bigg(\frac{\partial}{\partial\Bar{\xi}_\alpha}\frac{\partial}{\partial\lambda^a}+\Bar{\xi}^\alpha\frac{\partial}{\partial\lambda^a}\bigg).
\end{align}
The conformal $\mathfrak{conf}(2,  1)$ generators act on the super-currents in the usual way:
\begin{align}\label{conformalGenerators}
\mathcal{P}_{ab} 
&= \lambda_{(a}\bar{\lambda}_{b)},   
& \mathcal{K}_{ab} 
&= \frac{\partial^2}{\partial \lambda^{(a}\,  \partial \bar{\lambda}^{b)}},  
\notag\\[6pt]
\mathcal{M}_{ab} 
&= \frac{1}{2}
\left(
    \lambda_{(a}\frac{\partial}{\partial \lambda^{b)}}
    + 
    \bar{\lambda}_{(a}\frac{\partial}{\partial \bar{\lambda}^{b)}}
\right),  
&
\mathcal{D} 
&= \frac{1}{2}
\left(
    \lambda^{a}\frac{\partial}{\partial \lambda^{a}}
    +
    \bar{\lambda}^{a}\frac{\partial}{\partial \bar{\lambda}^{a}}
    + 2
\right).
\end{align}
Compared to $\mathcal{N}=1$ SUSY, we have the additional $R-$symmetry to consider. This is the symmetry of the equation,
\begin{align}
    \{\mathcal{Q}_a^\alpha,\mathcal{Q}_b^\beta\}=\mathcal{P}_{ab}\delta^{\alpha\beta}.
\end{align}
Under $\mathcal{Q}_a^\alpha\to R^\alpha_{\alpha'}Q_a^{a\alpha'}$ we find,
\begin{align}
     \{\mathcal{Q}_a^\alpha,\mathcal{Q}_b^\beta\}=\mathcal{P}_{ab}\delta^{\alpha\beta}\to  R_{\alpha'}^{\alpha}R_{\beta'}^{\beta}\{\mathcal{Q}_a^{\alpha'},\mathcal{Q}_b^{\beta'}\}=R_{\alpha'}^{\alpha}R_{\beta'}^{\beta}\mathcal{P}_{ab}\delta^{\alpha'\beta'}=\mathcal{P}_{ab}\delta^{\alpha\beta},
\end{align}
if $R_{\alpha'}^{\alpha}\delta^{\alpha'\beta'}R_{\beta'}^{\beta}=\delta^{\alpha\beta}$ which implies that $R\in \mathfrak{so}(\mathcal{N})$. Therefore, the $R-$symmetry group in three dimensions for $\mathcal{N}-$extended SUSY is $SO(\mathcal{N})$. The finite action of this symmetry operation on the Grassmann twistor variables is simply $\xi^\alpha\to R^{\alpha}_{\alpha'}\xi^{\alpha'}$ and similarly for $\Bar{\xi}^{\alpha}$. Although we shall not require it explicitly since we will apply the finite transformation, the infinitesimal version of the $R-$symmetry generator is,
\begin{align}
    \mathcal{R}_{\alpha\beta}(\xi)=\xi_{[\alpha}\frac{\partial}{\partial \xi^{\beta]}},~~\mathcal{R}_{\alpha\beta}(\Bar{\xi})=\Bar{\xi}_{[\alpha}\frac{\partial}{\partial \Bar{\xi}^{\beta]}}.
\end{align}
The conformal Ward identities for correlators that enjoy these symmetries read,
\begin{align}\label{WardId}
    \sum_{i=1}^{n}\mathcal{G}_i\mathbf\Psi_n^{h_1\cdots h_n}=0,  \mathcal{G}_i\in\{\mathcal{P}_{iab},  \mathcal{K}_{iab},  \mathcal{M}_{iab},  \mathcal{D}_{i},  h_i,  \mathcal{Q}_{ia}^\alpha,  \mathcal{S}_{ia}^\alpha,\mathcal{R}_{\alpha\beta}\}.
\end{align}
Equipped with this setup, we now proceed to define the Grassmannian for $\mathcal{N}-$extended SCFT$_3$.
\subsection{The $\mathcal{N}-$Extended Super Orthogonal Grassmannian}
We begin by defining the building blocks of this formalism before stating our main result. The orthogonal Grassmannian OGr$(n,2n)$ is the space of $n-$dimensional null-planes in $\mathbb{R}^{n,n}$. The metric on this space is,
\begin{align}\label{metric}
    Q=\begin{pmatrix}
        0_{n\times n}&\mathbb{I}_{n\times n}\\
        \mathbb{I}_{n\times n}&0_{n\times n}
    \end{pmatrix}.
\end{align}
We define $\Lambda$, which is a $2n\times 2$ vector formed out of the spinor helicity variables of all the operators, repackaging the kinematic data as follows:
\begin{align}\label{Lambda}
    \Lambda=\begin{pmatrix}
        \lambda_1^1&\lambda_1^2\\
        \vdots &\vdots\\
        \lambda_n^1&\lambda_n^2\\
        \Bar{\lambda}_1^1&\Bar{\lambda}_1^2\\
        \vdots&\vdots\\
        \Bar{\lambda}_n^1&\Bar{\lambda}_n^2
    \end{pmatrix}.
\end{align}
Similarly, we repackage the data of the Grassmann twistors $\xi^\alpha$ and $\Bar{\xi}^\alpha$ into the following set of $2^n$ distinct $2n\times \mathcal{N}$ Grassmann ``phase" space vectors.
\begin{align}\label{XiVectorNeq1}
    \Bar{\Xi}_\alpha^{--\cdots -}=\begin{pmatrix}
        \xi_{1\alpha}\\
        \xi_{2\alpha}\\
        \vdots\\
        \xi_{n\alpha}\\
        \frac{\partial}{\partial\xi_1^\alpha}\\
\frac{\partial}{\partial\xi_2^\alpha}\\
        \vdots\\
        \frac{\partial}{\partial \xi_n^\alpha}
    \end{pmatrix},  \Bar{\Xi}_\alpha^{+-\cdots -}=\begin{pmatrix}
        \frac{\partial}{\partial\Bar{\xi}_1^\alpha}\\
        \xi_{2\alpha}\\
        \vdots\\
        \xi_{n\alpha}\\
        \Bar{\xi}_{1\alpha}\\
\frac{\partial}{\partial \xi_2^{\alpha}}\\
        \vdots\\
        \frac{\partial}{\partial \xi_n^{\alpha}}
    \end{pmatrix},  \cdots,  \Bar{\Xi}_{\alpha}^{+\cdots +}=\begin{pmatrix}
        \frac{\partial}{\partial\Bar{\xi}_1^\alpha}\\
        \frac{\partial}{\partial\Bar{\xi}_2^\alpha}\\
        \vdots\\
        \frac{\partial}{\partial\Bar{\xi}_n^\alpha}\\
        \Bar{\xi}_{1\alpha}\\
        \Bar{\xi}_{2\alpha}\\
        \vdots\\
        \Bar{\xi}_{n\alpha}
    \end{pmatrix}.
\end{align}
Collectively, we denote these vectors as $\Bar{\Xi}_\alpha^{h_1\cdots h_n}$ to indicate which one we should use in a given helicity configuration. Essentially, we describe negative helicity (integer-spin) super-currents using $\xi$ variables and positive helicity super-currents by $\Bar{\xi}$. This is in line with the conventions of our companion $\mathcal{N}=1$ paper for integer spin super-currents.

These quantities are the square roots of the metric \eqref{metric},
\begin{align}
    \{\Bar{\Xi}_{\alpha}^{A,h_1\cdots h_n},\Bar{\Xi}_{\beta}^{B,h_1\cdots h_n}\}=\delta_{\alpha\beta}Q^{AB}
\end{align}
With these ingredients in hand, we define the $\mathcal{N}-$extended
orthogonal super-Grassmannian.
\begin{center}
\fbox{%
\begin{minipage}{0.99\textwidth}
\textbf{The Orthogonal Super-Grassmannian for $\mathcal{N}=2,3,4$ SCFT$_3$}
\footnotesize
\begin{align}\label{SuperGrassmannianN}
   &\mathbf{\Psi}_n^{h_1\cdots h_n}=\int \frac{d^{n\times 2n}C}{\text{Vol}(\mathbb{GL}(n))}\delta(C\cdot Q\cdot C^T)\delta(C\cdot\Lambda)\Bigg[\hat{\delta}(C\cdot\Bar{\Xi}^{h_1\cdots h_n}) \mathcal{F}^{h_1\cdots h_n}(C)+\hat{\mathcal{U}}(C,\Bar{\Xi}^{h_1\cdots h_n}) \mathcal{G}^{h_1\cdots h_n}(C)\Bigg],  
\end{align}
\normalsize
$\mathbf{\Psi}_n^{h_1\cdots h_n}$ is an n-point super-correlator involving integer spin super-currents. $\mathcal{F}^{h_1\cdots h_n}$ and $\mathcal{G}^{h_1\cdots h_n}$  are the functions of the $n\times n$ minors of $n\times 2n$ matrix $C$ and transform with a factor of $\text{Det}(G)^{-(n-3)-\mathcal{N}}$ under a $\mathbb{GL}(n)$ transformation. The quantities $\hat{\delta}$ and $\hat{\mathcal{U}}$ are Grassmann-valued objects that ensure supersymmetry.
The above Grassmannian integral trivializes all $10+4\mathcal{N}+\frac{\mathcal{N}(\mathcal{N}-1)}{2}=\frac{\mathcal{N}^2+7\mathcal{N}+20}{2}$ super-conformal Ward identities. 
\end{minipage}%
}
\end{center}

The quantities $\delta(C.Q.C^T)$ and $\delta(C.\Lambda)$ trivialize the conformal Ward identities as discussed in \cite{Arundine:2026fbr}. The Grassmann delta function $\hat{\delta}(C.\Bar{\Xi})$, generalizing the $\mathcal{N}=1$ version of our companion paper takes the form,
\begin{align}
    \hat{\delta}(C.\Bar{\Xi})=\int d^{n\times \mathcal{N}}\theta~e^{\theta^{i\alpha}C_{iA}\Bar{\Xi}^{A}_{\alpha}}=\int d^{n\times \mathcal{N}}\theta e^{\theta^{i\alpha}\eta_{i\alpha}},
\end{align}
where we have defined the variables,
\begin{align}
    \eta_{i\alpha}=C_{iA}\Bar{\Xi}^{A}_{\alpha}.
\end{align}
We have also suppressed the helicity labels on $\Bar{\Xi}$ for ease of notation. The $\eta_{i\alpha}$ obey the following algebra:
\begin{align}
    \{\eta_{i\alpha},\eta_{j\beta}\}=C_{iA}C_{jB}\{\Bar{\Xi}^A_\alpha,\Bar{\Xi}^B_\beta\}=C_{iA}Q^{AB}C_{jB}\delta_{\alpha\beta}=0~~\text{since}~C.Q.C^T=0.
\end{align}
Therefore, the $\eta_{i\alpha}$ behave like ordinary Grassmann numbers under the support of the orthogonality constraint. Thus, we can use the usual multi-variable Grassmann delta function definition, which is,
\begin{align}\label{GrassmannDeltaFunction}
    \hat{\delta}(C.\Bar{\Xi})=\hat{\delta}(\eta)=\prod_{\alpha=1}^{\mathcal{N}}\prod_{i=1}^{n}\eta_{i\alpha}.
\end{align}
Thus, the $\mathcal{N}-$extended Grassmann delta function is simply a product of $\mathcal{N}$ copies of the the $\mathcal{N}=1$ result indexed by $\alpha\in\{1,\cdots,\mathcal{N}\}$. Thus, we can make use of the $\mathcal{N}=1$ results to determine the higher supersymmetry Grassmann delta functions. 

The quantity $\mathcal{U}(C,\Bar{\Xi})$ is another super-conformal building block which is given by,
\begin{align}\label{Ublock}
    \mathcal{U}^{h_1\cdots h_n}(C,{\bar\Xi}^{h_1\cdots h_n}) =\prod_{i=1}^{n}\int d^\mathcal{N}{\chi}_i~e^{-\chi_i\cdot\Bar{\chi}_i}~\hat{\delta}(C.\Bar{\Xi}^{-h_1\cdots -h_n}),
\end{align}
where $\chi_i$ and its Fourier conjugate  $\bar\chi_i$ represents for $\xi$ or $\bar\xi$ as per the Grassmann phase space vector($\bar{\bar\Xi}$) used in the above equation. 

One can note that out of these two Grassmann quantities viz $\hat\delta(C\cdot\Xi^{h_1\cdots h_n})$ and $\int\hat\delta(C\cdot\Xi^{-h_1\cdots -h_n})$, are both Grassmann even for even point correlators. In fact, they are not linearly independent for even points. For example, it is easy to check explicitly that for two and four point functions,
\begin{align}\label{Utodelta24}
    \hat{\mathcal{U}}^{h_1\cdots h_4}(C,{\bar\Xi}^{h_1\cdots h_4})\propto \hat{\delta}(C.\Bar{\Xi}^{h_1\cdots h_4}).
\end{align}
Therefore, we can simply re-absorb the proportionality constant into $\mathcal{G}^{h_1\cdots h_n}(C)$, yielding a unique solution for $n=2,4$. Indeed, due to this fact, we will simply set $\mathcal{G}=0$ for two and four-point functions.
For odd point correlators, on the other hand, $\hat{\delta}$ is Grassmann odd whereas $\mathcal{\hat{U}}$ is Grassmann even. The $\mathcal{U}$ block will be required for three-point functions, as we shall see subsequently.

% This block will be important in three important for odd point correlation functions for which this is another independent building block as we will shall discuss now. In the $(-\cdots -)$ helicity we have,}
% \textcolor{red}{
% \begin{align}\label{deltatoU}
%     \mathcal{U}^{-\cdots -}(C,\Bar{\Xi}^{-\cdots -})=\prod_{i=1}^{n}\int d^\mathcal{N}\Bar{\xi}~e^{-\xi_i\cdot\Bar{\xi}_i}~\hat{\delta}(C.\Bar{\Xi}^{+\cdots +})=(?)^{\mathcal{N}}e^{M_{ij}\xi_i\cdot\xi_j},
% \end{align}
% where $M_{ij}$ is given by ??. This quantity is simply the Grassmann Fourier transform of $\hat{\delta}(C.\Bar{\Xi}^{+\cdots +})$ which is a function of $\Bar{\xi}_1,\Bar{\xi}_n$ to a function of $\xi_1,\cdots,\xi_n$ and thus will be a solution to the supersymmetric Ward identity by construction.

% \textcolor{red}{Similar expressions can be obtained in other helicities.  However, for $n=3$, these solutions are distinct. They take the form ??.}

Finally, the properties that $\mathcal{F}^{h_1\cdots h_n}$ and $\mathcal{G}^{h_1\cdots h_n}$ need to have in order to ensure that the integrand is $\mathbb{GL}(n)$ invariant and the correlator has the correct helicity with respect to each super-current are the following:
\begin{align}\label{GLnEQ}
    &\mathcal{F}^{h_1\cdots h_n}(GC)=\text{Det}(G)^{-(n-3)-\mathcal{N}}\mathcal{F}^{h_1\cdots h_n}(C),  ~G\in \mathbb{GL}(n),\notag\\
    &\mathcal{G}^{h_1\cdots h_n}(GC)=\text{Det}(G)^{-(n-3)-\mathcal{N}}\mathcal{G}^{h_1\cdots h_n}(C),  ~G\in \mathbb{GL}(n),  
\end{align}
and,  
\begin{align}\label{helicityEQ}
    &\mathcal{F}^{h_1\cdots h_n}(\rho\cdot C)= \frac{1}{\rho_1^{2h_1}\cdots\rho_n^{2h_n}}\mathcal{F}^{h_1\cdots h_n}(C),\notag\\
    &\mathcal{G}^{h_1\cdots h_n}(\rho\cdot C)= \frac{1}{\rho_1^{2h_1+\mathcal{N}\text{Sgn}(h_1)}\cdots\rho_n^{2h_n+\mathcal{N}\text{Sgn}(h_n)}}\mathcal{G}^{h_1\cdots h_n}(C),
\end{align}
where $\rho=\text{diag}(\rho_1,  \rho_2,  \cdots \rho_n,  \frac{1}{\rho_1},  \frac{1}{\rho_2},  \cdots,  \frac{1}{\rho_n})$ represents the independent little group scaling of each operator that is translated to the $C$ matrix from the scaling $\Lambda\to \rho\cdot \Lambda$ under a little group transformation $\{\lambda_1,  \cdots,  \lambda_n,  \Bar{\lambda}_1,  \cdots\Bar{\lambda}_n\}\to \{\rho_1\lambda_1,  \cdots,  \rho_n\lambda_n,  \frac{\Bar{\lambda}_1}{\rho_1},  \cdots,  \frac{\Bar{\lambda}_n}{\rho_n}\}$. Essentially,   the barred columns (first $n$ columns of the $C$ matrix) scale like $\Bar{\lambda}_i$ whereas the unbarred columns scale like $\lambda_i$ ~\cite{Arundine:2026fbr}.

We shall now show how \eqref{SuperGrassmannianN} automatically solves the super-conformal Ward identities \eqref{WardId}.
\subsection{Proof of Super-Conformal Invariance}
Invariance under the conformal generators such as $\mathcal{P}_{ab}$ and $\mathcal{K}_{ab}$ follows from the exact arguments used in the non-supersymmetric Grassmannian \cite{Arundine:2026fbr}. Thus we need to check invariance under $Q_a^\alpha,S_{a}^{\alpha}$ and $R_{\alpha\beta}$. We begin with the special super-conformal generator. We have (suppressing helicity indices)
\begin{align}
    &\sum_{i=1}^{n}\mathcal{S}_{ia}^{\alpha}\mathbf{\Psi}_n=\Bar{\Xi}^{A\alpha}\cdot\frac{\partial}{\partial\Lambda^{A a}}\int \frac{d^{n\times 2n}C}{\text{Vol}(\mathbb{GL}(n))}\delta(C.Q.C^T)\delta(C.\Lambda)\hat{\delta}(C.\Bar{\Xi})\mathcal{F}(C)\notag\\
    &=\Bar{\Xi}^{A\alpha}C_{jA}\int \frac{d^{n\times 2n}C}{\text{Vol}(\mathbb{GL}(n))}\delta(C.Q.C^T)\frac{\partial}{\partial (C_{jB}\Lambda^{Ba})}\delta(C.\Lambda)\hat{\delta}(C.\Bar{\Xi})\mathcal{F}(C)\notag\\
    &=\int \frac{d^{n\times 2n}C}{\text{Vol}(\mathbb{GL}(n))}\delta(C.Q.C^T)\frac{\partial}{\partial (C_{jB}\Lambda^{Ba})}\delta(C.\Lambda)\eta_j^\alpha\hat{\delta}(\eta)\mathcal{F}(C)=0,
\end{align}
 since $\eta_j^\alpha \hat{\delta}(\eta)=0$.
 As for the supersymmetry generator, we first note that $C\cdot \Lambda=0$ and $C\cdot Q\cdot C^T=0$ implies that $\Lambda^A_a=(C_{jB}Q^{AB})(P_{\Lambda})_a^j$. We then have,
 \begin{align}
     \sum_{i=1}^{n}\mathcal{Q}^\alpha_{ia}=\Lambda^A_a Q_{AB}\Bar{\Xi}^{B\alpha}=C_{jC}Q^{AC}(P_{\Lambda})^j_a Q_{AB}\Bar{\Xi}^{B\alpha}=(P_{\Lambda})^j_a C_{jB}\Bar{\Xi}^{B\alpha}=(P_{\Lambda})^j_a \eta^\alpha_j.
 \end{align}
 Thus, when acting on $\mathbf{\Psi}_n$ we obtain zero since $\eta_j^\alpha\hat{\delta}(\eta)=0$.

 All that remains to be checked is invariance under an $R-$symmetry transformation. Rather than checking the infinitesimal Ward identity, we will directly prove that \eqref{SuperGrassmannianN} is invariant under a finite $R-$symmetry transformation. The only quantities charged under this symmetry are the Grassmann twistor variables, which occur through $\hat{\delta}(C.\Bar{\Xi})$ in the Grassmannian. Thus, we have to prove that \eqref{GrassmannDeltaFunction} is invariant under $\eta_{i\alpha}\to R_{\alpha}^{\beta}\eta_{i\beta}$. We have,
 \begin{align}
 &\prod_{\alpha=1}^{\mathcal{N}}\prod_{i=1}^{n}\eta_{i\alpha}\to \prod_{\alpha=1}^{\mathcal{N}}\prod_{i=1}^{n}R_\alpha^\beta \eta_{i\beta}\propto \prod_{i=1}^{n}(R_{\alpha_1}^{1}\cdots R_{\alpha_\mathcal{N}}^\mathcal{N})\eta_i^{\alpha_1}\cdots \eta_i^{\alpha_\mathcal{N}}=\prod_{i=1}^{n}\frac{\epsilon_{\alpha_1\cdots \alpha_\mathcal{N}}}{(\mathcal{N})!}\text{Det}(R)\eta_i^{\alpha_1}\cdots \eta_i^{\alpha_\mathcal{N}}\notag\\
     &=\prod_{\alpha=1}^{\mathcal{N}}\prod_{i=1}^{n}\eta_{i\alpha},~~\text{since}~~\text{Det}(R)=1~\text{as}~R\in SO(\mathcal{N}).
 \end{align}
 This completes the proof of the super-conformal invariance of the first term of \eqref{SuperGrassmannianN}. As for the other one, we see using the definition \eqref{Ublock}, that the same arguments ensure its super-conformal invariance.
\section{$\mathcal{N}=2$ Super-Correlation functions}\label{sec:Neq2Gen}
 Having developed the formalism in the previous section, we now construct explicit examples of two, three, and four-point functions. We begin with the study of super-correlators in $\mathcal{N}=2$ theories. The super-currents of interest to us have the following component expansion\footnote{This expression can be obtained from the $U(1)$ basis expressions \cite{Jain:2023idr} by a simple change of basis $ \xi_i^1=\frac{\xi_i+ \omega_i}{4},~~\xi_i^2=-\frac{i}{4}\big(\xi_i- \omega_i\big)$. }:
 \begin{align}\label{Neq2Fields}
     &\mathbf{J}_s^{+}(\lambda,\Bar{\lambda},\Bar{\xi})=-\frac{\epsilon^{\alpha\beta}\Bar{\xi}_\alpha\Bar{\xi}_{\beta}}{2}J_s^{+}(\lambda,\Bar{\lambda})+\Bar{\xi}_{\alpha}J_{s+\frac{1}{2}}^{\alpha +}(\lambda,\Bar{\lambda})+ J_{s+1}^{+}(\lambda,\Bar{\lambda}),\notag\\
     &\mathbf{J}_s^{-}(\lambda,\Bar{\lambda},\xi)=\frac{\epsilon^{\alpha\beta}\xi_{\alpha}\xi_{\beta}}{2}J_s^{-}(\lambda,\Bar{\lambda})+\epsilon_{\alpha\beta}\xi^{\alpha}J_{s+\frac{1}{2}}^{\beta -}(\lambda,\Bar{\lambda})+J_{s+1}^{-}(\lambda,\Bar{\lambda}),
 \end{align}
 where $s\in\mathbb{Z}_{> 0}$. In general, these fields can carry internal symmetry indices such as adjoint indices of $SU(N)$. However, for $s=1$, the multiplet contains the stress tensor as the highest spin component, and thus in that case, the $s=1$ current is a $U(1)$ gauge field (the current dual to the gravi-photon field, for instance) and cannot carry any internal symmetry indices. We discuss the $s=0$ case, which is relevant for $\mathcal{N}=2$ super-Yang Mills theory, exclusively, in section \ref{sec:Neq2SYM}\footnote{For illustrative purposes, we stick to $s\in\mathbb{Z}_{>0}$ in the discussion on general CFTs. For super-scalars, at the level of three points, the three-point component scalar correlator is non-zero in a generic CFT. This will necessitate the use of both Grassmann building blocks $\hat{\delta}$ and $\hat{\mathcal{U}}$, unlike the spinning case that requires only the latter, as we shall see. It is, however, a simple matter to incorporate scalars into the analysis. In the next section \ref{sec:Neq2SYM}, the component scalar three-point function, however, is zero, and thus the results are natural extensions of the spinning case.}. 
 
 The component currents appearing in the above expansions are rescaled by $\frac{1}{p^{s-1}}$ to ensure the simple action of the SCT generator as in \eqref{conformalGenerators}.
\subsection{Two point functions}
Let us first evaluate the Grassmann delta function, focusing on the $(--)$ helicity configuration. As we mentioned earlier \eqref{Utodelta24}, only the $\hat{\delta}$ Grassmann building block is required at the level of two points, and thus we set $\mathcal{G}=0$. We find using \eqref{GrassmannDeltaFunction},
\begin{align}
    \hat{\delta}(C.\Bar{\Xi})=\bigg(\frac{(1\Bar{1})+(2\Bar{2})}{2}+(\Bar{1}\Bar{2})\xi_1^1 \xi_2^1\bigg)\bigg(\frac{(1\Bar{1})+(2\Bar{2})}{2}+(\Bar{1}\Bar{2})\xi_1^2 \xi_2^2\bigg),
\end{align}
where the superscripts on the Grassmann variables are the values of the $R-$symmetry $SO(2)$ indices. Expanding the above product and covariantizing results in,
\begin{align}
    &\hat{\delta}(C.\Bar{\Xi})=\bigg(\frac{(1\Bar{1})+(2\Bar{2})}{2}\bigg)^2+\bigg(\frac{(1\Bar{1})+(2\Bar{2})}{2}\bigg)(\Bar{1}\Bar{2})\xi_1^\alpha\xi_2^\alpha+\frac{(\Bar{1}\Bar{2})^2}{2}(\xi_1^\alpha\xi_2^\alpha)^2\notag\\&=\bigg(\frac{(1\Bar{1})+(2\Bar{2})}{2}\bigg)^2\text{exp}\bigg({\frac{2(\Bar{1}\Bar{2})}{(1\Bar{1})+(2\Bar{2})}}\xi_1\cdot\xi_2\bigg)=(1\Bar{1})^2\text{exp}\bigg(\frac{(\Bar{1}\Bar{2})}{(1\Bar{1})}\xi_1\cdot \xi_2\bigg),
\end{align}
where we used the fact that $(1\Bar{1})=(2\Bar{2})$ in the right branch, which is where the two-point functions are supported.
The super-Grassmannian is thus given by,
\footnotesize
\begin{align}\label{Neq2Psi2}
    \Psi_2^{-s,-s}=\int \frac{d^{2\times 4}C}{\text{Vol}(\mathbb{GL}(2))}\delta(C.Q.C^T)\delta(C.\Lambda)(1\Bar{1})^2\text{exp}\bigg(\frac{(\Bar{1}\Bar{2})}{(1\Bar{1})}\xi_1\cdot \xi_2\bigg)\mathcal{F}^{-s,-s}(C).
\end{align}
\normalsize
Satisfying $\mathbb{GL}(2)$ invariance of the integrand and the correct helicity scaling leads us to the following ansatz for $\mathcal{F}^{-s,-s}(C)$:
\begin{align}\label{Neq2F2pt}
    \mathcal{F}^{+s,+s}(C)=\frac{(12)^{2(s+1)}}{(1\Bar{1})^{2s+3}}.
\end{align}
We now substitute \eqref{Neq2F2pt} in \eqref{Neq2Psi2}, gauge fix and use $C.Q.C^T$ to obtain,
\begin{align}\label{2pointRightBranch}
    C=\begin{pmatrix}
        1&0&0&-c_{12}\\
        0&1&c_{12}&0
    \end{pmatrix}.
\end{align}
This results in,
\begin{align}
    \mathbf{\Psi}_2^{-s,-s}&=\int dc_{12}\delta^2(\lambda_1-c_{12}\Bar{\lambda}_2)\delta^2(\lambda_2+c_{12}\Bar{\lambda}_1)\text{exp}\bigg(\frac{\xi_1\cdot \xi_2}{c_{12}}\bigg)c_{12}^{2s+1}\notag\\
    &\propto \frac{\langle 1 2\rangle^{2(s+1)}}{E^{2(s+1)-1}}\bigg(1-\frac{\langle\Bar{1}\Bar{2}\rangle}{E}\xi_1^\alpha\xi_2^\alpha-\frac{\langle\Bar{1}\Bar{2}\rangle^2}{2E^2}(\xi_1^\alpha\xi_2^\alpha)^2\bigg),
\end{align}
where we have suppressed the overall momentum-conserving delta function. By comparing with the super-field expansion, one finds the correct conformally invariant component two-point functions. To compare with the results of \cite{Jain:2023idr}, we switch to the $U(1)$ basis from the $SO(2)$ one we are currently using as follows:
\begin{align}
    \xi_i^1=\frac{\xi_i+ \omega_i}{4},~~\xi_i^2=-\frac{i}{4}\big(\xi_i- \omega_i\big).
\end{align}
In this basis we find (also using $E=p_1+p_2=2p_1$),
\begin{align}
    \mathbf{\Psi}_{2}^{+s,+s}\propto \frac{\langle 12 \rangle^{2s}}{p_1^{2s-1}}\bigg(\frac{\xi_{1+}\omega_{1+}\xi_{2+}\omega_{2+}}{16}-\frac{\langle 1 2\rangle}{4p_1}(\omega_{1+}\xi_{2+}+\xi_{1+}\omega_{2+})+\frac{\langle 1 2\rangle^2}{p_1^2}\bigg).
\end{align}
The term in the parentheses matches the building block found in \cite{Jain:2023idr} perfectly, thus serving as a check of our new formalism. Since the analysis for the $(++)$ helicity two-point function is identical, we do not present the details here.

\subsection{Three point functions}
Next, we consider three-point functions, focusing on the $(+++)$ helicity configuration for illustrative purposes. Given the super-field \eqref{Neq2Fields} that we have, we observe that only the $\hat{\mathcal{U}}(C,\bar{\bar\Xi})$ Grassmann block is going to have a non-trivial contribution. For example, setting $s=1$ results in a top component $\langle JJJ\rangle$, which is zero for the Abelian case. Hence, $\hat{\mathcal{U}}(C,\bar{\bar\Xi})$ using \eqref{Ublock} for the $(+++)$ super-correlator is given by,
            \begin{align}
                 \hat{\mathcal{U}}^{+++}(C)&= \prod_{i=1}^{3}\int d^2\xi_i^\alpha~~ e^{-\xi_i \cdot\bar\xi_i} ~ \hat\delta(C\cdot({\bar\Xi}^{\beta})^{---}) \notag \\ 
                 &  = -\bigg(\prod_{i=1}^{i=3}\int d\xi_i^{(1)} e^{-\xi^{(1)}_i\bar\xi^{(1)}_i} \hat\delta(C\cdot({\bar\Xi}^1)^{---}) \bigg)\bigg(\prod_{i=1}^{i=3}\int d\xi_j^{(2)} e^{-\xi^{(2)}_j \bar\xi^{(2)}_j} \hat\delta(C\cdot({\bar\Xi}^2)^{---})\bigg) \notag\\
                 &=  -  4(\bar1\bar2\bar3)^2 \text{exp}\bigg[-\sum_{i,j,k,l=1}^3{\epsilon_{ijk}\bar\xi_i\cdot\bar\xi_j \frac{(\bar k l\bar l)}{(\bar1\bar2\bar3)}} \bigg], 
            \end{align}
where repeated indices are summed over from 1 to 3. It is now a simple matter to use the Grassmannian integral \eqref{SuperGrassmannianN} to determine the function $\mathcal{G}$ by matching with the $\order{1}$ term, which is the correlator $\langle TTT\rangle$. Its expression is \cite{Arundine:2026fbr},
  \begin{align}
        \mathcal{G}^{+,+,+}(C) = \frac{(\Bar{1}\Bar{2}\Bar{3})^2}{((1\Bar{1}2)(\Bar{2}3\Bar{3}))^2}.
      \end{align}

\subsection{Four point functions}
We now proceed to $n=4$ points. As discussed earlier in equation \eqref{Utodelta24}, only the $\hat{\delta}(c.\Bar{\Xi})$ block is relevant at four points since $\mathcal{U}(C,\Bar{\Xi})$ is proportional to it. Thus, we set $\mathcal{G}=0$ in the super-Grassmannian \eqref{SuperGrassmannianN}. Also, we focus on the $(-+-+)$ helicity configuration, although a similar analysis can be repeated for other helicities. The $\mathcal{N}=2$ Grassmann delta function is simply a product of two copies of its $\mathcal{N}=1$ counterpart. For $\alpha=1,2$ we find (no sum over $\alpha$),
\footnotesize
\begin{align}
    &\hat{\delta}(C.(\Bar{\Xi}^{\alpha})^{-+-+})=\frac{S+T-U}{2}\Bigg[1+\frac{1}{S+T-U}\bigg(\xi_1^\alpha\Bar{\xi}_2^\alpha\big((\Bar{1}\Bar{3}23)-(\Bar{1}\Bar{4}24)\big)+\xi_1^\alpha\xi_3^\alpha\big((\Bar{1}\Bar{3}\Bar{4}4)-(\Bar{1}\Bar{2}\Bar{3}2)\big)\notag\\&+\xi_1^\alpha\Bar{\xi}_4^\alpha\big((\Bar{1}\Bar{2}24)-(\Bar{1}\Bar{3}34)\big)+\Bar{\xi}_2^\alpha\xi_3^\alpha\big((\Bar{3}\Bar{4}24)-(\Bar{1}\Bar{3}12)\big)+\Bar{\xi}_2^\alpha\Bar{\xi}_4^\alpha\big((\Bar{3}234)-(\Bar{1}124)\big)+\xi_3^\alpha\Bar{\xi}_4^\alpha\big((\Bar{1}\Bar{3}14)-(\Bar{2}\Bar{3}24)\big)\notag\\&+2(\Bar{1}2\Bar{3}4)\xi_1^\alpha\Bar{\xi}_2^\alpha\xi_3^\alpha\Bar{\xi}_4^\alpha\bigg)\Bigg]\notag\\&=\frac{S+T-U}{2}\text{exp}\Bigg(\frac{2\bigg(\xi_1^\alpha\Bar{\xi_2}^{\alpha}(\Bar{1}\Bar{3}23)+\xi_1^\alpha\xi_3^\alpha(\Bar{1}\Bar{3}\Bar{4}4)+\xi_1^\alpha\Bar{\xi}_4^\alpha(\Bar{1}\Bar{2}24)+\Bar{\xi}_2^\alpha\xi_3^\alpha(\Bar{3}\Bar{4}24)+\Bar{\xi}_2^\alpha\Bar{\xi}_4^\alpha(\Bar{3}234)+\xi_3^\alpha\Bar{\xi}_4^\alpha(\Bar{1}\Bar{3}14)\bigg)}{(S+T-U)}\Bigg).
\end{align}
\normalsize
Taking a product of the above quantity with $\alpha=1$ and $\alpha=2$ yields,
\footnotesize
\begin{align}\label{deltaCXi4pointNeq2}
    &\hat{\delta}(C.\Bar{\Xi}^{-+-+})=\hat{\delta}(C.(\Bar{\Xi}^{1})^{-+-+})\hat{\delta}(C.(\Bar{\Xi}^{2})^{-+-+})\notag\\
    &=\bigg(\frac{S+T-U}{2}\Bigg)^2\text{exp}\Bigg(\frac{2\bigg(\xi_1\cdot\Bar{\xi}_2(\Bar{1}\Bar{3}23)+\xi_1\cdot\xi_3(\Bar{1}\Bar{3}\Bar{4}4)+\xi_1\cdot \xi_4(\Bar{1}\Bar{2}24)+\Bar{\xi}_2\cdot\xi_3(\Bar{3}\Bar{4}24)+\Bar{\xi}_2\cdot\Bar{\xi}_4(\Bar{3}234)+\xi_3\cdot\Bar{\xi}_4(\Bar{1}\Bar{3}14)\bigg)}{(S+T-U)}\bigg).
\end{align}
\normalsize
The next step is to bootstrap $\mathcal{F}^{-+-+}(C)$ that appears in the super-Grassmannian \eqref{SuperGrassmannianN}. Using the super-field expansion, we find that using \eqref{Neq2Fields} the coefficient of the highest order term which in component form is $\xi_{1}^{1}\xi_1^{2}\Bar{\xi}_2^{1}\Bar{\xi}_2^{2}\xi_3^1\xi_3^2\Bar{\xi}_4^1\Bar{\xi}_4^2$ is $\psi^{-s_1,+s_2,-s_3,+s_4}$. Using this information along with \eqref{deltaCXi4pointNeq2} which gives a factor of $4(\Bar{1}2\Bar{3}4)T^2$ that multiplies this Grassmann structure and using the Grassmannian \eqref{SuperGrassmannianN}, we find,
\begin{align}\label{Neq2fourpointconstraint}
    \mathcal{F}^{-s_1,+s_2,-s_3,+s_4}(C)=\frac{A^{-s_1,+s_2,-s_3,s_4}}{4(\Bar{1}2\Bar{3}4)^2}.
\end{align}
$A^{-s_1,+s_2,-s_3,s_4}$ is the Grassmann space correlator corresponding to $\psi^{-s_1,+s_2,-s_3,+s_4}$. Using this information, every other component correlator is determined in terms of just the bottom component. Thus, the general four-point function is given by,
\footnotesize
\begin{align}\label{fourpointgenresult}
    &\mathbf{\Psi}^{-s_1,+s_2,-s_3,+s_4}\notag\\&=\int \frac{d^{4\times 8}C}{\text{Vol}(\mathbb{GL}(4))}\delta(C.Q.C^T)\delta(C.\Lambda)\bigg(\frac{S+T-U}{2}\Bigg)^2\notag\\
    &\times\text{exp}\Bigg(\frac{2\bigg(\xi_1\cdot\Bar{\xi}_2(\Bar{1}\Bar{3}23)+\xi_1\cdot\xi_3(\Bar{1}\Bar{3}\Bar{4}4)+\xi_1\cdot \xi_4(\Bar{1}\Bar{2}24)+\Bar{\xi}_2\cdot\xi_3(\Bar{3}\Bar{4}24)+\Bar{\xi}_2\cdot\Bar{\xi}_4(\Bar{3}234)+\xi_3\cdot\Bar{\xi}_4(\Bar{1}\Bar{3}14)\bigg)}{(S+T-U)}\bigg)\frac{A^{-s_1,+s_2,-s_3,s_4}}{4(\Bar{1}2\Bar{3}4)^2}.
\end{align}
\normalsize
We now proceed to test this general formula in the particular and interesting holographic setting of $\mathcal{N}=2$ super Yang-Mills theory in AdS$_4$.
\section{AdS$_4$ $\mathcal{N}=2$ SYM}\label{sec:Neq2SYM}
The subject of this section is AdS$_4$ super Yang-Mills theory. We first define the super-fields, discuss the constraints from supersymmetry in the Grassmannian framework, and use the scalar four-point function to bootstrap the gluon MHV correlator. 
\subsection{The $\mathcal{N}=2$ Super-Gluon Field}
The spectrum of $\mathcal{N}=2$ SYM theory consists of two scalars $O$ and $\Bar{O}$ which are CPT conjugates, four fermions $O_{\frac{1}{2}}^{\alpha\pm}$ and two gluons $J^{\pm}$, all of which are in the adjoint representation of $SU(N)$. These degrees of freedom are packaged into the following super-fields.
 \begin{align}\label{Neq2SYMFields}
      &\mathbf{J}_0^{+}(\lambda,\Bar{\lambda},\Bar{\xi})=-\frac{\epsilon^{\alpha\beta}\Bar{\xi}_\alpha\Bar{\xi}_{\beta}}{2}O(\lambda,\Bar{\lambda})+\Bar{\xi}_{\alpha}O_{\frac{1}{2}}^{\alpha +}(\lambda,\Bar{\lambda})+ J^{+}(\lambda,\Bar{\lambda}),\notag\\
     &\mathbf{J}_0^{-}(\lambda,\Bar{\lambda},\xi)=\frac{\epsilon^{\alpha\beta}\xi_{\alpha}\xi_{\beta}}{2}\Bar{O}(\lambda,\Bar{\lambda})+\epsilon_{\alpha\beta}\xi^{\alpha}O_{\frac{1}{2}}^{\beta -}(\lambda,\Bar{\lambda})+J^{-}(\lambda,\Bar{\lambda}),
 \end{align}
This is exactly the same structure as in flat space \cite{Elvang:2011fx}, which is expected since the theory is classically conformally invariant. Using supersymmetry, we have found \eqref{Neq2fourpointconstraint} which relates every component correlator to just one. For \eqref{Neq2SYMFields} in the $(-+-+)$ helicity configuration, we will use the correlator $\langle \Bar{O}O\Bar{O}O\rangle$ as the seed. First, we proceed to bootstrap this correlator and then use supersymmetry to derive the spin$-\frac{1}{2}$ and spin$-1$ expressions.

\subsection{Bootstrapping the Scalar four point function}
The scalar four-point function in this theory can receive contributions from a gluon exchange as well as a quartic self-interaction. In \cite{Arundine:2026fbr}, the expression for charged scalars exchanging a photon was derived. In the non-abelian case, with a gluon exchange, the kinematic expression remains the same as in \cite{Arundine:2026fbr}. In the colour ordering of interest to us where the $s$ and $t$ channels contribute, we should simply add the contributions. Allowing for a quartic interaction term with an arbitrary coefficient results in the following ansatz for the scalar four-point function:
\begin{align}\label{Neq2scalar4pt}
    A^{\Bar{O},O,\Bar{O},O}=\frac{1}{S+T+U}\frac{T-U}{S}+\frac{1}{S+T+U}\frac{S-U}{T}+\frac{\lambda_4}{S+T+U},
\end{align}
 We will fix $\lambda_4$ by demanding that this results in the correct gluino four-point function that we have computed in our companion paper.
\subsection{Deriving the gluino and gluon four-point functions}
We find using the super-Grassmannian \eqref{fourpointgenresult} (setting $A^{-s_1,+s_2,-s_3,+s_4}=A^{\Bar{O},O,\Bar{O},O}$),
\begin{align}\label{FcNeq2SYM}
    \mathcal{F}^{\Bar{O},O,\Bar{O},O}=\frac{A^{\Bar{O},O,\Bar{O},O}}{4(\Bar{1}2\Bar{3}4)^2}=\frac{1}{4(\Bar{1}2\Bar{3}4)^2}\bigg(\frac{1}{S+T+U}\frac{T-U}{S}+\frac{1}{S+T+U}\frac{S-U}{T}+\frac{\lambda_4}{S+T+U}\bigg).
\end{align}
It is now a simple matter to use this in the super-Grassmannian \eqref{SuperGrassmannianN} (with $\mathcal{G}=0$) and the super-field expansion \eqref{Neq2SYMFields} to determine the remaining component correlators. For example, by looking at the coefficient of $\xi_1^1\Bar{\xi}_2^1\xi_3^1\Bar{\xi}_4^1$ we should find the $(-+-+)$ gluino four point function. Using \eqref{FcNeq2SYM} results in,
\begin{align}
    A^{-\frac{1}{2},\frac{1}{2},-\frac{1}{2},\frac{1}{2}}=\frac{-(S+T-U)}{8(\Bar{1}2\Bar{3}4)^2}\bigg(\frac{1}{S+T+U}\frac{T-U}{S}+\frac{1}{S+T+U}\frac{S-U}{T}+\frac{\lambda_4}{S+T+U}\bigg).
\end{align}
The expected result obtained from the analysis in our companion paper is,
\begin{align}
    A^{-\frac{1}{2},\frac{1}{2},-\frac{1}{2},\frac{1}{2}}|_{\text{expected}}=\frac{-(1\Bar{2}3\Bar{4})}{2(S+T+U)}\bigg(\frac{1}{S}+\frac{1}{T}\bigg).
\end{align}
Subtracting the expected answer from the one obtained from SUSY and using,
\begin{align}
    (\Bar{1}2\Bar{3}4)(1\Bar{2}3\Bar{4})=\bigg(\frac{S+T-U}{2}\bigg)^2,
\end{align}
results in,
\begin{align}
    \frac{(\lambda_4-2)(1\Bar{2}3\Bar{4})}{2(S+T-U)(S+T+U)}=0\implies \lambda_4=2.
\end{align}
Therefore, having fixed the quartic coefficient in the scalar four-point function \eqref{Neq2scalar4pt} we find,
\begin{align}\label{Neq2scalar4ptfixed}
       A^{\Bar{O},O,\Bar{O},O}&=\frac{1}{S+T+U}\frac{T-U}{S}+\frac{1}{S+T+U}\frac{S-U}{T}+\frac{2}{S+T+U}\notag\\
       &=\frac{S+T-U}{S+T+U} \left(\frac{1}{S}+\frac{1}{T}\right).
\end{align}
As another sanity check, we look at the coefficient of the $\order{1}$ term in the Grassmann expansion, which is the $(-+-+)$ MHV gluon correlator. Using the updated scalar four point function \eqref{Neq2scalar4ptfixed} in \eqref{FcNeq2SYM} yields,
\begin{align}
    A^{-1,+1,-1,+1}&=\frac{-(S+T-U)^2}{4
    (\Bar{1}2\Bar{3}4)^2}\Bigg(\frac{1}{S+T+U}\frac{T-U}{S}+\frac{1}{S+T+U}\frac{S-U}{T}+\frac{2}{S+T+U}\Bigg)\notag\\
    &=-2\bigg(\frac{1}{S+T+U}+\frac{1}{S+T-U}\bigg)\frac{(1\Bar{2}3\Bar{4})^2}{ST},
\end{align}
which perfectly matches the correct result \cite{Arundine:2026fbr}. Therefore, using $\mathcal{N}=2$ SUSY, we are able to bootstrap the gluon correlator given its partner scalar counterpart. 

Thus, we obtain,
\footnotesize
\begin{align}\label{fourpointNeq2SYM}
    &\mathbf{\Psi}^{-s_1,+s_2,-s_3,+s_4}\notag\\&=\int \frac{d^{4\times 8}C}{\text{Vol}(\mathbb{GL}(4))}\delta(C.Q.C^T)\delta(C.\Lambda)\frac{1}{(S+T+U)(S+T-U)}\left(\frac{1}{S}+\frac{1}{T}\right)(1\Bar{2}3\Bar{4})^2\notag\\
    &\times\text{exp}\Bigg(\frac{2\bigg(\xi_1\cdot\Bar{\xi}_2(\Bar{1}\Bar{3}23)+\xi_1\cdot\xi_3(\Bar{1}\Bar{3}\Bar{4}4)+\xi_1\cdot \xi_4(\Bar{1}\Bar{2}24)+\Bar{\xi}_2\cdot\xi_3(\Bar{3}\Bar{4}24)+\Bar{\xi}_2\cdot\Bar{\xi}_4(\Bar{3}234)+\xi_3\cdot\Bar{\xi}_4(\Bar{1}\Bar{3}14)\bigg)}{(S+T-U)}\bigg).
\end{align}
\normalsize

\section{$\mathcal{N}=4$ Super-Correlators and AdS$_4$ SYM}\label{Neq4SYM}
One can work out the general CFT construction given our formalism, however for illustrative purposes, we choose to work with the application to AdS. The aim of this section is to apply the results of the Grassmannian Construction to N=4 super Yang-Mills theory in AdS$_4$ with gauge group SU(N). The spectrum in bulk consists of gluons, scalars, and fermions in the adjoint representation of SU(N). $\mathcal{N}=4$ super Yang-Mills theory is a classically conformally invariant theory, so its spectrum is the same as in AdS$_4$ and flat space-time. The supersymmetric action in the flat space and AdS will be similar, modulo boundary terms, due to conformal equivalence.

We will focus on the boundary-to-boundary supercorrelator and an essential step before proceeding is to choose the supermultiplet to work with. We start the first discussion with the construction of the CPT self-conjugate super-field containing s= 0,$\frac{1}{2},1$ operators. This is followed by a discussion of two-point functions using this super-field and then we construct the four-point function, finding a remarkably simple expression that encapsulates all the gluon, gluino, and scalar super-correlators in every helicity configuration. We then construct a super-operator whose components range from spin-zero up to spin-two, and compute its two-point function to illustrate the effectiveness of the formalism in a general CFT setting. 

\subsection{CPT Self-Conjugate Superfield in AdS$_4$ $\mathcal{N}=4$ SYM}
The bosonic spectrum of $\mathcal{N}=4$ super Yang-Mills theory consists of two gluons (positive and negative helicity) and six scalars and eight fermionic degrees of freedom, consisting of four species of fermions, each with a positive and negative helicity component. Taking inspiration from the on-shell super-field in four-dimensional flat space \cite{Elvang:2011fx}, we package these degrees of freedom into the following CPT self-conjugate super-conformal multiplet.
% \begin{equation}
%     \mathbf{J}_s = J_{s-2} + \frac{1}{3!} \epsilon_{\alpha \beta\gamma\delta} {\xi}^\alpha \lambda_{s-\frac{3}{2}}^{\beta\gamma\delta} + \frac{1}{2!} \frac{1}{2!} \epsilon_{\alpha\beta\gamma\delta} {\xi}^\alpha {\xi}^\beta S^{\gamma\delta}_{s-1} + \frac{1}{3!} \epsilon_{\alpha\beta\gamma\delta} {\xi}^\alpha {\xi}^\beta {\xi}^\gamma \lambda^\delta_{s-\frac{1}{2}} + \frac{1}{4!} \epsilon_{\alpha\beta\gamma\delta}~{\xi}^\alpha {\xi}^\beta {\xi}^\gamma {\xi}^\delta J_{s} .
% \end{equation}
\begin{equation}\label{eq:N4SYMsuper-field}
    \mathbf{J} = J^{+}_1 + \bar\xi^\alpha J_{\frac{1}{2}}^{+\alpha} -\frac{1}{2!} \bar\xi^\alpha \bar\xi^\beta J_0^{\alpha\beta} - \frac{1}{3!} \bar\xi^{\alpha}\bar\xi^{\beta}\bar\xi^{\gamma} J_{\frac{1}{2}}^{-\alpha\beta\gamma} + \bar\xi^1\bar\xi^2 \bar\xi^3 \bar\xi^4 J_1^-
\end{equation}
where \(J_1^{\pm}\) are the positive and negative spin-1 currents, \(J_{\frac{1}{2}}^{+\alpha}\) and \(J_{\frac{1}{2}}^{-\alpha\beta\gamma}\) are eight fermions dual to each other, and \(J_0^{\alpha\beta}\) are the six scalars. 

Being a self-CPT conjugate multiplet requires the condition,
   \begin{align}\label{CPTconjgscalar}
      \bar J_{0~\alpha\beta} = -\frac{1}{2!} \epsilon_{\alpha\beta\gamma\delta} J_0^{\gamma\delta} .   
   \end{align}
We will start by bootstrapping the supercorrelators using the above supermultiplet.
\subsubsection{Two point function}
The super-field is CPT self-conjugate, and to compute its two-point function, we choose the building block $\hat\delta(C\cdot{\bar\Xi}^{++})$ which is given by,
\begin{align}\label{eq:N42ptDelta}
\hat\delta(C\cdot \bar\Xi) = \notag& \frac{1}{16}\Big(16 \bar\xi_1^{1}\bar\xi_1^{2}\bar\xi_1^{3} \bar\xi_1^{4} \bar\xi_2^{1}\bar\xi_2^{2}\bar\xi_2^{3} \bar\xi_2^{4} ~(12)^4  \\\notag &+ 8\frac{1}{(3!)^2} \epsilon_{\alpha\beta\gamma\delta}\epsilon_{\rho\sigma\tau\delta} \bar\xi_1^{\alpha}\bar\xi_1^{\beta}\bar\xi_1^{\gamma}\bar\xi_2^{\rho}\bar\xi_2^{\sigma}\bar\xi_2^{\tau} ~(12)^3 ((1\bar{1}) + (2\bar{2})) \\
\notag& -  4\frac{1}{(2!)^3} \epsilon_{\alpha\beta\gamma\delta}\epsilon_{\rho\sigma\gamma\delta} \bar\xi_1^{\alpha}\bar\xi_1^{\beta}\bar\xi_2^{\rho}\bar\xi_2^{\sigma} ~(12)^2 ((1\bar{1}) + (2\bar{2}))^2 \\\notag&- 2\bar\xi_1^{\alpha}\bar\xi_2^{\alpha} ~(12) ((1\bar{1}) + (2\bar{2}))^3 \\&+ ((1\bar{1}) + (2\bar{2}))^4\Big).
\end{align}
 We bootstrap the \(\mathcal{F}(C)\) using the principles of section \ref{sec:Formalism} to be 
\begin{equation}
    \mathcal{F}(C) = \frac{(\bar 1\bar 2)^{2}}{(1\bar 1)^5}.
\end{equation}
Now, performing the Grassmannian integral \eqref{SuperGrassmannianN} (with $\mathcal{G}=0$ at two points due to it being redundant \eqref{Utodelta24}) to obtain the two-point supercorrelator as: 
\begin{align}
\mathbf{\Psi}_2 = -\Big(\notag& \bar\xi_1^{1}\bar\xi_1^{2}\bar\xi_1^{3} \bar\xi_1^{4} \bar\xi_2^{1}\bar\xi_2^{2}\bar\xi_2^{3} \bar\xi_2^{4} ~\frac{\langle 12 \rangle^2}{E}  - \frac{1}{(3!)^2} \epsilon_{\alpha\beta\gamma\delta}\epsilon_{\rho\sigma\tau\delta} \bar\xi_1^{\alpha}\bar\xi_1^{\beta}\bar\xi_1^{\gamma}\bar\xi_2^{\rho}\bar\xi_2^{\sigma}\bar\xi_2^{\tau} ~\langle 12 \rangle \\
& -  \frac{1}{(2!)^3} \epsilon_{\alpha\beta\gamma\delta}\epsilon_{\rho\sigma\gamma\delta} \bar\xi_1^{\alpha}\bar\xi_1^{\beta}\bar\xi_2^{\rho}\bar\xi_2^{\sigma} ~E + \bar\xi_1^{\alpha}\bar\xi_2^{\alpha} ~ \frac{E^2}{\langle 12 \rangle} + \frac{E^3}{\langle 12 \rangle^2}\Big).
\end{align}
Using \(\frac{E}{\langle12\rangle} = \frac{\langle \bar 1\bar 2\rangle}{E}\), we can write this as 
\begin{align}
\mathbf{\Psi}_2 = -\Big(\notag& \bar\xi_1^{1}\bar\xi_1^{2}\bar\xi_1^{3} \bar\xi_1^{4} \bar\xi_2^{1}\bar\xi_2^{2}\bar\xi_2^{3} \bar\xi_2^{4} ~\frac{\langle 12 \rangle^2}{E}  - \frac{1}{(3!)^2} \epsilon_{\alpha\beta\gamma\delta}\epsilon_{\rho\sigma\tau\delta} \bar\xi_1^{\alpha}\bar\xi_1^{\beta}\bar\xi_1^{\gamma}\bar\xi_2^{\rho}\bar\xi_2^{\sigma}\bar\xi_2^{\tau} ~\langle 12 \rangle \\
& -  \frac{1}{(2!)^3} \epsilon_{\alpha\beta\gamma\delta}\epsilon_{\rho\sigma\gamma\delta} \bar\xi_1^{\alpha}\bar\xi_1^{\beta}\bar\xi_2^{\rho}\bar\xi_2^{\sigma} ~E + \bar\xi_1^{\alpha}\bar\xi_2^{\alpha} ~E\langle \bar1\bar2 \rangle + \frac{\langle \bar1\bar2 \rangle^2}{E}\Big).
\end{align}
This result matches the known CFT results not only for \(\langle J^- J^-\rangle\) but also \(\langle J_{\frac{1}{2}}J_{\frac{1}{2}}\rangle\) and  \(\langle J_0 J_0 \rangle\), and therefore confirms the validity of our formalism. 

An analysis similar to that in section \ref{sec:Neq2Gen} can be carried out for three-point functions where the $\hat{\mathcal{U}}$ block features, and we do not present the details here. Our main interest, however, is in the four-point case, which we now turn to.

\subsubsection{Four point function}
In this subsection, we compute the four-point supercorrelator using the spin-1 four-point correlator as input. 
The \(\mathcal{N}=1\) \(\hat\delta_1(C\cdot\bar\Xi^{++++})\) for four-point can be written abstractly as 
\begin{equation}
M_0 \exp\left (\sum_{i<j} \frac{M_{ij}}{M_0} \bar\xi_i \bar\xi_j \right) 
\end{equation}
where
\begin{equation}\label{eq:N4fourptdeltaMs}
    M_0 = \frac{1}{2}(S+T+U), ~M_{ij} =\frac{1}{2}\sum_{k\ne i,j}( i j\bar k k)
\end{equation}
from which we can construct the \(\mathcal{N}=4\) \(\hat\delta(C\cdot\bar\Xi)\) as 
\begin{equation}\label{eq:N4fourptdelta}
    \hat\delta(C\cdot \bar \Xi) = M_0^4 \prod_{\alpha = 1}^4\exp\left (\sum_{i<j} \frac{M_{ij}}{M_0} \bar\xi_i^\alpha \bar\xi_j^\alpha \right) = M_0^4 \exp\left( \sum_{i<j} \frac{M_{ij}}{M_0} \bar\xi_i \cdot \bar\xi_j \right)
\end{equation}
where the dot product \(\bar\xi_i \cdot \bar\xi_j = \delta_{\alpha\beta} \bar\xi_i^\alpha \bar\xi_j^\beta\), and we could write the product of the exponentials as an exponent of the sum since \(\bar\xi_i\cdot \bar\xi_j\) is Grassmann even and commutes. \\
We bootstrap the \(\mathcal{F}(C)\) by demanding that the component without \(\bar\xi\) should go over to \(A^{+1,+1,+1,+1}\), and we obtain it to be 
\begin{equation}
    \mathcal{F}(C) = \frac{32 (\bar1\bar2\bar3\bar4)^2}{ST (S+T+U)^4}\left( \frac{1}{S+T+U} + \frac{1}{S+T-U}  \right)
\end{equation}
Therefore, the supercorrelator in Grassmann space is just
\begin{align}
&\mathbf{\Psi}_4=\int \frac{d^{4\times 8}C}{\text{Vol}(\mathbb{GL}(4))}\delta(C.Q.C^T)\delta(C.\Lambda)\notag\\
    &~~~~~~~~~~~~~~~~~~~~~~~ 2 \exp\left( \sum_{i<j} \frac{M_{ij}}{M_0} \bar\xi_i \cdot \bar\xi_j \right)\frac{ (\bar1\bar2\bar3\bar4)^2}{ST}\left( \frac{1}{S+T+U} + \frac{1}{S+T-U}  \right),
\end{align}
which is an elegant and simple result, almost as much as its flat space counterpart. 

With the above supercorrelator, we can check whether it reproduces some of the known correlators in Grassmannian. 

To get the gluon four-point function in a different helicity $A^{-1,-1,-1,-1}$ using the above supercorrelator expression, we look at the coefficient of $\prod_{i=1}^4\bar\xi_i^1\bar\xi_i^2\bar\xi_i^3\bar\xi_i^4$ which is: 
    \begin{align}
        \langle J_1^-J_1^-J_1^-J_1^-\rangle = \frac{(1234)^2}{(S+T+U)^4} \bigg(\frac{1}{S+T+U}+ \frac{1}{S+T-U}\bigg), 
    \end{align}
which is the correct expected gluon four-point function. 

Let us check the scalar 4 point function by looking at the coefficient of $\bar\xi_1^1\bar\xi_1^2\bar\xi_2^3\bar\xi_2^4\bar\xi_3^1\bar\xi_3^2\bar\xi_4^3\bar\xi_4^4$ which is given as:
    \begin{align}
        \langle J_0^{12}J_0^{34}J_0^{12}J_0^{34}\rangle &= \frac{4(S+T)((\bar1\bar32\bar2)+(\bar1\bar34\bar4))^2((\bar2\bar41\bar1)+(\bar2\bar4 3\bar3))^2(1234)^2}{ST(S+T-U)(S+T+U)^5} \notag \\
        & = \frac{1}{S+T+U}\frac{T-U}{S} + \frac{1}{S+T+U}\frac{S-U}{T} + \frac{2}{S+T+U},
    \end{align}
When going from the first line to the second, we have used Pl\"ucker relations. We are getting the correct scalar four-point function that we obtained in $\mathcal{N}=2$   in eq \eqref{Neq2scalar4ptfixed} with gluon exchange and the contact diagram. 
Now, let us try to understand why this combination of scalar correlator four-point function has these exchanges. For this, we should look at the interaction term in the bulk/flat Lagrangian\cite{Elvang:2013cua} for the scalar and the gluon interaction: 
    \begin{align}
        L_{int} \sim \int Tr(D_{\mu}\phi^{\alpha\beta}D^{\mu}\phi_{\alpha\beta}) + Tr([\phi^{\alpha\beta},\phi^{\gamma\delta}][\phi_{\alpha\beta},\phi_{\gamma\delta}]). 
    \end{align}
The interaction term is schematically given by $\phi^{\alpha\beta}\phi_{\alpha\beta} A $, $\phi^{\alpha\beta}\phi_{\alpha\beta} A A$ and $\phi^{\alpha\beta}\phi^{\gamma\delta}\phi_{\alpha\beta}\phi_{\gamma\delta}$ . Hence, the four-point scalar colour ordered amplitude using eq \eqref{CPTconjgscalar}
     \begin{align}
         \langle\phi^{12}\phi_{12}\phi^{12}\phi_{12}\rangle = \langle\phi^{12}\bar\phi^{34}\phi^{12}\bar\phi^{34}\rangle.
     \end{align}
will contain the gluon exchange s, t channel and contact diagram as per the interaction allowed. This is the exact correlator we are computing on the boundary $\langle J_0^{12}J_0^{34}J_0^{12}J_0^{34}\rangle$.

   It is a simple exercise to obtain spin half four point function using the above results. This spin-half correlator, depending on the choices of R-index, can get contributions from gluon exchange or also can arise solely from scalar exchange via Yukawa couplings \cite{Elvang:2013cua}. For the gluon exchange part, the result should be identical to what is reported in section \ref{sec:Neq2SYM}. Our formalism naturally captures both these contributions, highlighting an interesting difference over the $\mathcal{N}=2$ framework.
\subsection{Stress tensor two-point function using \(\mathcal{N}=4\) SUSY}
We can extend the construction in \eqref{eq:N4SYMsuper-field} to other spins, at the expense of losing the CPT-self-conjugacy. Specifically, for \(s=2\) super-field, we construct the positive helicity super-field to be 
\begin{equation}\label{eq:N4Graviton}
    \mathbf{J}^+_2 = J_2^+ + \bar\xi^\alpha J_{\frac{3}{2}}^{+\alpha} -\frac{1}{2!} \bar\xi^\alpha \bar\xi^\beta J_{1}^{+\alpha\beta} - \frac{1}{3!} \bar\xi^{\alpha}\bar\xi^{\beta}\bar\xi^{\gamma}J_{\frac{1}{2}}^{+\alpha\beta\gamma} + \bar\xi^1\bar\xi^2 \bar\xi^3 \bar\xi^4 ~J_0
\end{equation}
where now \(J_0\) is a scalar, \(J_2^+\) is the positive helicity Graviton, \(\lambda\) and \(J\)s are eight fermions, and \(J\) are six spin-1 currents in positive helicity. 
For \(++\) helicity, the same \(\hat\delta(C\cdot\Xi)\) as in \eqref{eq:N42ptDelta} can be carried over, and we bootstrap the \(\mathcal{F}(C)\) to be 
\begin{equation}
    \mathcal{F}^{+2, +2}(C) = \frac{(\bar 1\bar 2)^4}{(1\bar 1)^7}
\end{equation}
which gives the supercorrelator to be 
\begin{align}
\mathbf{\Psi}^{+2,+2}_2 \notag&= -\Big( \bar\xi_1^{1}\bar\xi_1^{2}\bar\xi_1^{3} \bar\xi_1^{4} \bar\xi_2^{1}\bar\xi_2^{2}\bar\xi_2^{3} \bar\xi_2^{4} ~E - \frac{1}{(3!)^2} \epsilon_{\alpha\beta\gamma\delta}\epsilon_{\rho\sigma\tau\delta} \bar\xi_1^{\alpha}\bar\xi_1^{\beta}\bar\xi_1^{\gamma}\bar\xi_2^{\rho}\bar\xi_2^{\sigma}\bar\xi_2^{\tau} ~\frac{E^2}{\langle 12 \rangle} \\\notag
&~~~~~~~~~~~~~~~~~ -  \frac{1}{(2!)^3} \epsilon_{\alpha\beta\gamma\delta}\epsilon_{\rho\sigma\gamma\delta} \bar\xi_1^{\alpha}\bar\xi_1^{\beta}\bar\xi_2^{\rho}\bar\xi_2^{\sigma} ~\frac{E^3}{\langle 12 \rangle^2} + \bar\xi_1^{\alpha}\bar\xi_2^{\alpha} ~\frac{E^4}{\langle 12 \rangle^3} + \frac{E^5}{\langle 12 \rangle^4}\Big)\\\notag 
& = -\Big( \bar\xi_1^{1}\bar\xi_1^{2}\bar\xi_1^{3} \bar\xi_1^{4} \bar\xi_2^{1}\bar\xi_2^{2}\bar\xi_2^{3} \bar\xi_2^{4} ~E - \frac{1}{(3!)^2} \epsilon_{\alpha\beta\gamma\delta}\epsilon_{\rho\sigma\tau\delta} \bar\xi_1^{\alpha}\bar\xi_1^{\beta}\bar\xi_1^{\gamma}\bar\xi_2^{\rho}\bar\xi_2^{\sigma}\bar\xi_2^{\tau} ~E \langle \bar1\bar2 \rangle \\
&~~~~~~~~~~~~~~~~~ -  \frac{1}{(2!)^3} \epsilon_{\alpha\beta\gamma\delta}\epsilon_{\rho\sigma\gamma\delta} \bar\xi_1^{\alpha}\bar\xi_1^{\beta}\bar\xi_2^{\rho}\bar\xi_2^{\sigma}~E \langle \bar1\bar2 \rangle^2 + \bar\xi_1^{\alpha}\bar\xi_2^{\alpha}~E \langle \bar1\bar2 \rangle^3 + ~E \langle \bar1\bar2 \rangle^4\Big)
\end{align}
This supercorrelator gives the correct known results of \(\langle SS\rangle\) correlator and \(\langle TT\rangle\)
The analysis for \(--\) helicity follows similarly.

This construction is particularly useful for obtaining the stress-tensor four-point function starting from the scalar four-point correlator. In the AdS$_4$ context, this suggests that for $\mathcal{N}=4$ supergravity, the graviton four-point function can be reconstructed from the scalar correlator. We plan to report this result in a subsequent work.
\section{The Flat Space Limit}\label{sec:FlatLimit}
In this section, we take the flat space limit of our CFT$_3$/AdS$_4$ results for $\mathcal{N}=2$ and $\mathcal{N}=4$ SYM theories. Our goal is to match with the known elegant results of \cite{Elvang:2011fx}, especially in $\mathcal{N}=4$, and also to see explicitly the $R-$symmetry enhancement from $SO(\mathcal{N})$ to $SU(\mathcal{N})$. 

To obtain the flat space scattering amplitude from the Grassmannian integral,  the first step is to gauge-fix the $\mathbb{GL}(n)$ redundancy of the $C$ matrix. In the right branch that will be relevant for the calculations to follow, one can choose $C$ to take the following form:
\begin{align}\label{rightBranchGaugeFixing}
    C=\begin{pmatrix}
        1&0&0&0&0&-c_{12}&-c_{13}&-c_{14}\\
        0&1&0&0&c_{12}&0&-c_{23}&-c_{24}\\
        0&0&1&0&c_{13}&c_{23}&0&-c_{34}\\
        0&0&0&1&c_{14}&c_{24}&c_{34}&0
    \end{pmatrix}.
\end{align}
Using $\delta(C.\Lambda)$, we can solve for $5$ out of the $6$ Schwinger parameters $c_{ij}$. Parametrizing them following \cite{Arundine:2026fbr} results in,
\begin{align}
    c_{ij}=\frac{\langle i j\rangle}{E}+\frac{\tau}{2}\epsilon_{ijkl}\langle \Bar{k}\Bar{l}\rangle, 
\end{align}
where $E=p_1+p_2+p_3+p_4$ is the total energy. The flat space limit corresponds to taking the residue of the Grassmannian integral at $\tau=0$ and taking the $E\to 0$ limit.

\subsection{$\mathcal{N}=2$}
We begin by taking the flat limit of the four-point function in $\mathcal{N}=2$ SYM theory \eqref{fourpointNeq2SYM}. When taking the flat limit, we gauge fix and parameterize the elements of the $C$ matrix as discussed above. When taking the $E\to 0$ limit, we find that the form-factor $\frac{(S+T)(1\Bar{2}3\Bar{4})^2}{ST(S+T+U)(S+T-U)}\propto \frac{E^3}{\tau}$. Therefore, we need to focus on the term from the exponent in \eqref{fourpointNeq2SYM} that is independent of $\tau$ and goes like $\frac{1}{E^4}$ in order to obtain a non-zero residue at $\tau=0$ and has the correct exponent as $E\to 0$. Performing this analysis results in,
\begin{align}
    A^{-+-+}=\frac{\langle 13\rangle^2}{\langle 12\rangle\langle 23\rangle\langle 34\rangle\langle 4 1\rangle}\prod_{\alpha=1}^{2}\Bigg(\langle 13\rangle-\langle 12\rangle\xi_3^\alpha\Bar{\xi}_2^\alpha-\langle 14\rangle \xi_3^\alpha\Bar{\xi}_4^\alpha-\langle 2 3\rangle\xi_1^\alpha\Bar{\xi}_2^\alpha+\langle 34\rangle\xi_1^\alpha\Bar{\xi}_4^\alpha-\langle 2 4\rangle\xi_1^\alpha\xi_3^\alpha\Bar{\xi}_2^\alpha\Bar{\xi}_4^\alpha\bigg).
\end{align}
Comparing to the known flat space results \cite{Elvang:2011fx} (see equation $(2.16)$ and truncate to $\mathcal{N}=2$), we find that by setting $\xi_i^\alpha=-\Bar{\eta}_i^\alpha$ and $\Bar{\xi}_i^\alpha=\eta_i^\alpha$ reveals a perfect match. 

\subsection{$\mathcal{N}=4$ SYM}
The \(\mathcal{N}=4\) \(\hat\delta(C\cdot \Xi)\) \eqref{eq:N4fourptdelta} can be expanded as 
\begin{equation}
    \hat\delta(C\cdot \Xi) = \prod_\alpha \left( M_0 + \left(\sum_{i<j} M_{ij} \bar\xi_i^\alpha \bar\xi_j^\alpha \right) +\frac{1}{2}\frac{1}{M_0}\left(\sum_{i<j}M_{ij} \bar\xi_i^\alpha \bar\xi_j^\alpha \right)^2~  \right),
\end{equation}
where the \(M_{ij}\)s and \(M_0\) are 
\begin{equation}
    M_0 = \frac{1}{2}(S+T+U), ~M_{ij} = \frac{1}{2}\sum_{k\ne i,j}( i j\bar k k)
\end{equation}
To evaluate the residue at \(\tau  = 0\), we use the parameterization as in the previous subsection, and in this parameterization, we see that 
\begin{equation}
    M_0 = -E \tau ,~~M_{ij} = \tau\langle i j \rangle  + \frac{E}{2}\tau^2 \epsilon_{ijkl}\langle \bar k\bar l \rangle.
\end{equation} 
Further, if one expands the quadratic term and uses Pl\"ucker relations, one will discover that the entire term simply becomes
\begin{equation}
     \frac{1}{2}\frac{1}{M_0}\left(\sum_{i<j}M_{ij} \bar\xi_i^\alpha \bar\xi_j^\alpha \right) ^2 = (1234) \bar\xi_1^\alpha \bar\xi_2^\alpha \bar\xi_3^\alpha \bar\xi_4^\alpha = \tau^2 E^2 ~ \bar\xi_1^\alpha \bar\xi_2^\alpha \bar\xi_3^\alpha \bar\xi_4^\alpha.
\end{equation} 
Let us note that we are working with the parameterization such that  \((1234) = \tau^2 E^2\).\\
Therefore, we have 
\begin{equation}
    \hat\delta(C\cdot\Xi) = \prod_\alpha \left(-E\tau + \left(\sum_{i<j} \left( \tau\langle i j \rangle  +\frac{1}{2} E\tau^2 \epsilon_{ijkl}\langle \bar k\bar l \rangle\right) \bar\xi_i^\alpha \bar\xi_j^\alpha \right) +\tau^2 E^2 \bar\xi_1^\alpha \bar\xi_2^\alpha \bar\xi_3^\alpha \bar\xi_4^\alpha  \right).
\end{equation}
The \(\mathcal{F}(C)\) is
\begin{equation}
        \mathcal{F}(C) = \frac{64 (S+T) (\bar1\bar2\bar3\bar4)^2}{ST (S+T-U)(S+T+U)^5},
\end{equation}
which evaluates to
\begin{equation}
    \frac{1}{E^4 \tau^4}\left(\frac{2} {ST }\frac{1}{(S+T-U)} - \frac{1} {ST }\frac{1}{E\tau}\right).
\end{equation}
Now, we can take the \(\displaystyle\frac{1}{E^4 \tau^4}\) into the product, and get 
\begin{equation}
       \hat\delta(C\cdot \Xi) \mathcal{F}(C) =  \prod_\alpha \left( 1 + \sum_{i<j} \left( \frac{1}{E}\langle i j \rangle  + \frac{\tau}{2} \epsilon_{ijkl}\langle \bar k\bar l \rangle\right) \bar\xi_i^\alpha \bar\xi_j^\alpha  +\tau E \bar\xi_1^\alpha \bar\xi_2^\alpha \bar\xi_3^\alpha \bar\xi_4^\alpha \right)\left(\frac{2} {ST }\frac{1}{(S+T-U)} - \frac{1} {ST }\frac{1}{E\tau}\right).
\end{equation}
Since in the \(\hat\delta(C\cdot\Xi)\) there is no \(\tau = 0\) pole, the first term from \(\mathcal{F}\) does not contribute anything to the residue at \(\tau=0\). Further, the terms that are proportional to \(\tau\) inside the exponent will also produce extra \(\tau\)s in the numerator when expanded, and therefore will not contribute to the residue at all. The only contribution to the residue will come from 
\begin{equation}
    -\prod_\alpha\left(  1 + \sum_{i<j}  \frac{1}{E}\langle i j \rangle  \bar\xi_i^\alpha \bar\xi_j^\alpha \right)\left(\frac{1} {ST }\frac{1}{E\tau}\right).
\end{equation}
We have 
\begin{equation}
     \frac{1} {ST }\frac{1}{E}  \bigg|_{\tau = 0}=  \frac{1} {\langle \bar 1 \bar 2 \rangle \langle \bar 3 \bar 4 \rangle \langle \bar 1 \bar 4 \rangle \langle \bar 2 \bar 3 \rangle\tau_s \bar\tau_s \tau_t \bar\tau_t }  =  \frac{E^3}{\langle  1  2 \rangle \langle  2  3 \rangle \langle  3  4 \rangle\langle  1  4 \rangle}.
\end{equation}
Now one can see that the leading singularity in \(E\) will come from only  \(\displaystyle \prod_\alpha\left(\sum_{i<j}  \frac{1}{E}\langle i j \rangle  \bar\xi_i^\alpha \bar\xi_j^\alpha \right)\) since any terms in the product involving \(1\) will be subleading in \(E=0\) singularity, and therefore we do not need to consider those terms in the flat limit. \\
The flat limit, therefore, is
\begin{equation}
     A_4 = \frac{-1}{\langle  1  2 \rangle \langle  2  3 \rangle \langle  3  4 \rangle\langle  1  4 \rangle} \frac{1}{2^4}\prod_\alpha\left(\sum_{i,j}  \langle i j \rangle  \bar\xi_i^\alpha \bar\xi_j^\alpha \right).
\end{equation}
We see that this structure exactly matches the known flat space scattering amplitude \cite{Elvang:2011fx} (see equation $(2.8)$) perfectly.
\begin{equation}
    A_4 = \frac{1}{\langle  1  2 \rangle \langle  2  3 \rangle \langle  3  4 \rangle\langle  1  4 \rangle} \delta^{(8)}(\tilde Q),~~~~\delta^{(8)}(\tilde Q) = \displaystyle\frac{1}{2^4}\prod_{\alpha}\sum_{i,j} \langle ij\rangle\bar\xi_i^\alpha \bar\xi_j^\alpha.
\end{equation}
Again, we see that the terms that do not respect the \(SU(4)_R\) symmetry but are allowed by the \(SO(4)_R\) symmetry drop out in the flat limit. The symmetry enhancement from \(SO(4)_R\) in in SCFT\(_3\) to \(SU(4)_R\) in the scattering amplitudes of \(\mathcal N = 4\) SYM is explicitly demonstrated here.

\section{Discussion}\label{sec:Discussion}
In this work, we have developed a Super-Grassmannian framework for $n$-point functions in $\mathcal{N}=2$ to $\mathcal{N}=4$ SCFT$_3$, extending our earlier construction for $\mathcal{N}=1$. A central feature of this formalism is that super-conformal symmetry and $R$-symmetry are implemented in a manifest and unified manner, leading to a set of purely algebraic relations among correlators within the same multiplet. This provides a significant simplification compared to approaches based on spinor helicity variables, where supersymmetry typically leads to a mixture of algebraic and differential constraints \cite{Bala:2026new1}.

We have demonstrated the utility of this framework through explicit examples in AdS$_4$ super Yang--Mills theories. In particular, we showed that nontrivial spinning observables can be reconstructed starting from simpler scalar correlators. In the $\mathcal{N}=4$ case, we have established a direct flat-space limit that reproduces known $\mathcal{N}=4$ SYM amplitudes, offering a nontrivial check of the formalism and highlighting its connection to familiar amplitude structures.

Our results open several interesting directions for future work. A natural next step is the systematic study of higher-spin observables, in particular the four-graviton correlator, where the Grassmannian approach may provide a more efficient route compared to traditional bootstrap methods. It would also be important to further clarify the flat space limit of our construction, especially in the $\mathcal{N}=4$ case. In particular, it is natural to ask whether structures familiar from flat-space amplitudes—such as dual conformal invariance
and Yangian symmetry—have a direct counterpart in the CFT language, or whether they emerge only after taking an appropriate flat-space limit. Understanding how (or if) these symmetries are encoded in the correlator framework could shed light on the deeper geometric principles underlying conformal/AdS correlators.

We should point out that the four-point graviton correlator has not been computed in this work, and its analysis within our framework remains an important open problem. It would also be interesting to extend our construction to Vasiliev theory. In particular, given recent progress on four-point scalar correlators \cite{De:2026shn}, it would be worthwhile to investigate how spinning correlators are captured in this formalism.

%More broadly, our construction suggests the existence of an underlying geometric picture for supersymmetric correlators in three dimensions, analogous to the role played by Grassmannians and related structures in flat-space scattering amplitudes. It would be interesting to explore whether this perspective can be developed further, potentially leading to new geometric formulations or dual descriptions of CFT observables. We hope that the present work provides a useful step in this direction.

\section*{\Large Acknowledgment}
AB acknowledges a UGC-JRF fellowship. AAR acknowledges a CSIR-JRF fellowship.\\ D K.S. would like to thank Saurabh Pant for many discussions over the years on (supersymmetric) scattering amplitudes.
\newpage
\appendix

\section{Notation and Conventions}\label{app:Notation}
In this appendix, we state our notations and conventions. For spacetime $SL(2)$ spinor indices we use lower-case latin alphabets such as $a,b,c,\cdots\in\{1,2\}$. The letters $i,j\in\{1,\cdots,n\}$, however, are reserved for the particle index. Lower-case Greek alphabets $\alpha,\beta\cdots\in\{1,\cdots,\mathcal{N}\}$ are used to denote $SO(\mathcal{N})$ $R-$symmetry vector indices. These indices are raised and lowered using $\delta^{\alpha\beta}$. We denote a dot product between two Grassmann twistors $\xi_i^{\alpha}$ and $\xi_j^\beta$ as $\xi_i\cdot\xi_j=\xi_i^\alpha\xi_j^\beta\delta^{\alpha\beta}$. 

Next, we discuss our notation regarding the Grassmannian. Consider the little group scaling viz $\{\lambda_1, \cdots, \lambda_n, \Bar{\lambda}_1, \cdots\Bar{\lambda}_n\}\to \{\rho_1\lambda_1, \cdots, \rho_n\lambda_n, \frac{\Bar{\lambda}_1}{\rho_1}, \cdots, \frac{\Bar{\lambda}_n}{\rho_n}\}$.  For \(C\cdot \Lambda\) to be little group invariant,  the \(C\) matrix should transform in the following way:  $\{c_1, \cdots, c_n, {c}_{n+1}, ~\cdots {c}_{2n}\}\to \{\frac{c_1}{\rho_1}, \cdots, \frac{c_n}{\rho_n},  ~\rho_1 {c}_{n+1}, \cdots, \rho_n {c}_{2n}\}$. Thus,  we label the \(C\) matrix as,
 \begin{equation}
     \{c_1, \cdots, c_n, {c}_{n+1}, ~\cdots {c}_{2n}\} \equiv \{ \bar 1,  \cdots \bar n,  1, \cdots n \}
 \end{equation}
 where in this notation, the quantity \((i_1 i_2\cdots i_n)\)  \(i\) can involve entries that are either barred or unbarred and are short for,
 \begin{equation}
     (i_1 i_2\cdots i_n) \equiv \det(\{i_1,  i_2, \cdots i_n  \}).
 \end{equation}
More details on the geometry of the Grassmannian and specific gauge choices to evaluate the integrals can be found in Appendix $B$ of our companion paper \cite{Bala:2026new1}. 

%\textcolor{red}{I think that the material in the main-text is enough.}

\bibliography{biblio}
\bibliographystyle{JHEP}
\end{document}